\documentclass[twocolumn, amsmath,amssymb,amssymb,aps, showpacs, prl]{revtex4}
\usepackage{amssymb}
\usepackage{amsmath}
\usepackage{graphicx}
\usepackage{dcolumn}
\usepackage{bm}
\usepackage{braket}
\usepackage{mathrsfs}
\usepackage{graphicx}
\begin{document}
\title{Quantum coherence-assisted propagation of surface plasmon polaritons}
\author {Pankaj K. Jha$^{1}$, Xiaobo Yin$^{1,2}$ and Xiang Zhang$^{1,2,}$}
\affiliation{$^{1}$NSF Nanoscale Science and Engineering Center(NSEC), 5130 Etcheverry Hall, University of California, Berkeley, California 94720, USA\\
$^{2}$Materials Science Division, Lawrence Berkeley National Laboratory, 1 Cyclotron Road Berkeley, California 94720, USA
 }
\date{\today}
\pacs{240.6680, 130.2790.}    
\begin{abstract}
We theoretically demonstrate coherent control over propagation of surface plasmon polaritons(SPP), at both telecommunication and visible wavelengths,  on a metallic surface adjacent to quantum coherence (phaseonium) medium composed of three-level quantum emitters (semiconductor quantum dots, atoms, rare-earth ions, etc.) embedded in a dielectric host. The coherent drive allows us to provide sufficient gain for lossless SPP propagation and also lowers the pumping requirements.  In case of lossy propagation, an order of magnitude enhancement in propagation length can be achieved. Optical control over SPP propagation dynamics via an external coherent drive holds promise for quantum control in the field of nanophotonics. 
\end{abstract}
\maketitle

Sub-wavelength confinement of electromagnetic radiation, by coupling it to free electrons in metals, has led to the development of sensing nanoscale molecular complexes\cite{Voronine12} even down to single molecule\cite{Sharma12}, surface plasmon-polariton (SPP) based lasers\cite{Berini12}, ultrafast processing of optical signals\cite{StockOE} etc. to name a few. On the other hand quantum coherence and interference effects in atomic and molecular physics have been extensively studied due to its intriguing counterintuitive physics and potential important applications\cite{ScullyZub}. Extending coherence effects  to plasmonics is often encountered with sever challenges like ultrafast (1-10 fs) relaxation time scale of the surface plasmons (SP) and large intrinsic losses\cite{AgraMills}. These road blocks limit the realization of SPP based practical optical devices.  

Amplification of localized SP and SPP using gain medium like quantum dots(QDs) has gained interests due to its ability to compensate the energy dissipation limits\cite{spaser1,Seidel05,Noginov09,Oulton2009,Grand09}. Unfortunately the gain provided by active medium is not always sufficient due to impractical requirements \cite{Khurgin12a,Khurgin12b} or competing processes like amplified spontaneous emission of SPP(ASESPP) which may limit the gain available for loss compensation\cite{Bolger12}. 

In this Letter, we enhance the propagation length of SPPs, which depends on the internal and radiation damping\cite{AgraMills},  via quantum coherence. We consider SPP propagation on a planar metallic surface adjacent to quantum coherence (phaseonium) medium\cite{Scully92} composed of three-level quantum emitters (semiconductor quantum dots, atoms, rare-earth ions, molecules, etc.) embedded in a dielectric host as shown in Fig.1. Three-level systems experience Fano-type interference in their absorption profile that generates an asymmetry between absorption and stimulated emission. Here we apply this asymmetry to mitigate the SPPs absorption, thus reducing the radiative damping of SPPs. It is worth mentioning here that such asymmetry may lead to lasing without inversion(LWI)\cite{LWI}. We demonstrate that propagation of SPP with large intrinsic losses can benefit from quantum boost using coherent drive which can act as an external control parameter.
\begin{figure}[t]
\centerline{\includegraphics[height=4.3cm,width=0.44\textwidth,angle=0]{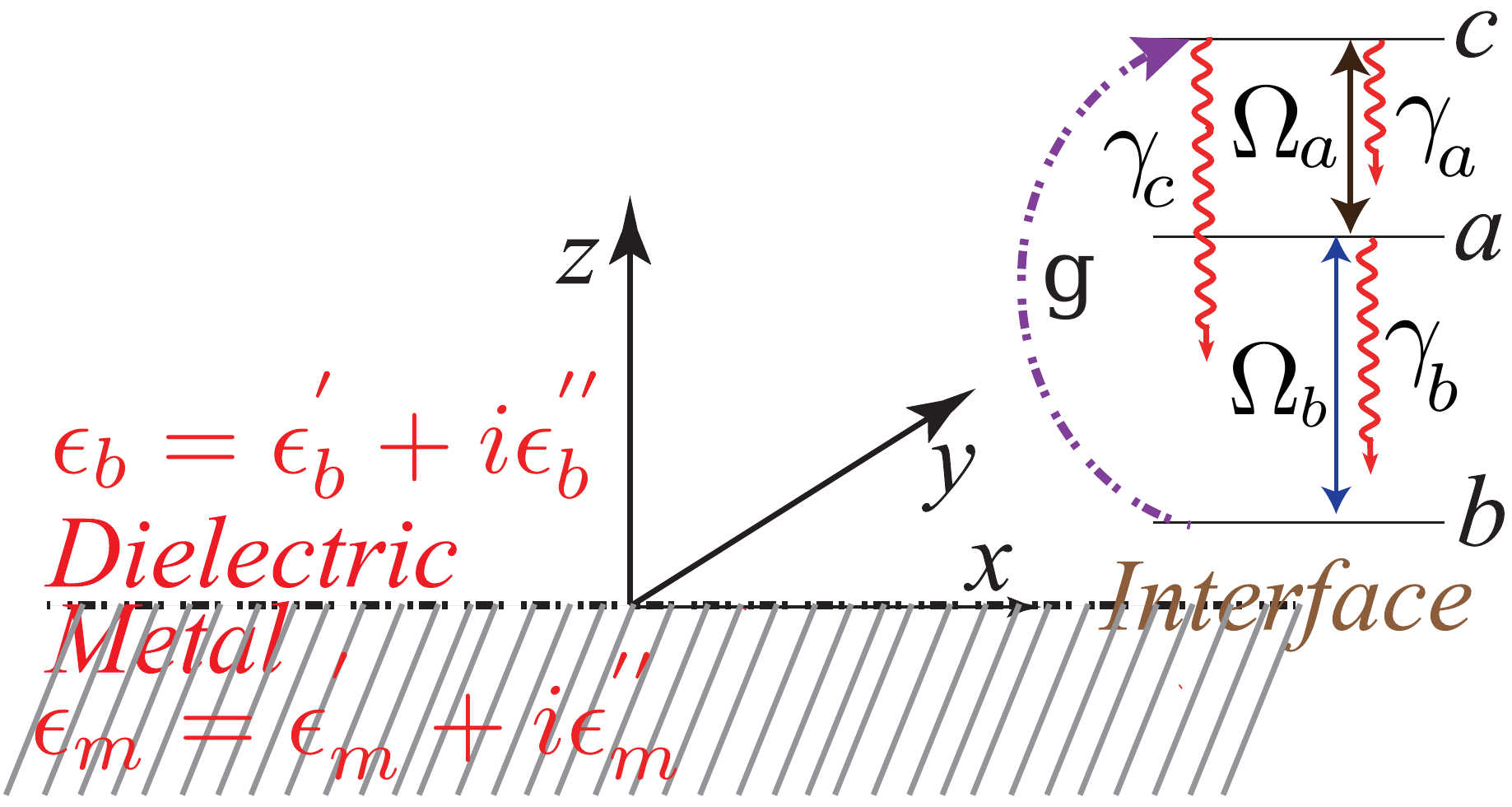}}
\caption{Schematic of the quantum coherence assisted propagation of surface plasmon polariton on a metallic waveguide adjacent to quantum coherence (phaseonium) medium composed of three-level emitters (semiconductor quantum dots, atoms, rare-earth ions, molecules, etc.) embedded in a dielectric host. The gain medium is incoherently pumped (optical or electrical) at a rate $g$ to the upper level $|c\rangle$ which can decay to levels $|a\rangle$ and $|b\rangle$.  We select the three-level emitter such that the $|a\rangle \leftrightarrow |b\rangle$ transition is resonantly coupled (near-field dipole-coupling) to the plasmon mode of the metal surface and the emission from this transition is efficiently transferred to SPPs.}
\label{SPPFig1}
\end{figure}

In our model we assume SPPs propagating along the positive $x$ direction on a metal-phaseonium(MP) interface lying in the $x$-$y$ plane. The TM waves can be written as,
\begin{equation}\label{fields1}
\begin{split}
\textbf{E}_{\alpha}=(1/2)(\mathcal{E}_{x},0,\mathcal{E}_{z,\alpha})\exp[i(k_{x}x+k_{z,\alpha}z-\nu t)]
 \end{split}
 \end{equation} 
 \begin{equation}\label{fields2}
\begin{split}
 \textbf{H}_{\alpha}=(1/2)(0,\mathcal{H}_{z,\alpha},0)\exp[i(k_{x}x+k_{z,\alpha}z-\nu t)],
 \end{split}
 \end{equation}
 where the indices $\alpha=(b,m)$ denote the phaseonium and the metal regions respectively. From the Maxwell's equations and continuity at the boundary one can readily obtain the SPP dispersion relations\cite{AgraMills}
\begin{equation}\label{eq1}
\begin{split}
k^{2}_{x}=k^{2}_{0}\left(\frac{\epsilon_{b}\epsilon_{m}}{\epsilon_{b}+\epsilon_{m}}\right), \quad k^{2}_{z,\alpha}=k^{2}_{0}\left(\frac{\epsilon^{2}_{\alpha}}{\epsilon_{b}+\epsilon_{m}}\right).
\end{split}
\end{equation}
Here $k_{0}=\omega/c$ is the free space wave vector of the incident radiation. The imaginary part of $k_{x}$ characterizes the SPP field attenuation during its propagation along the metal-phaseonium(MP) interface. The gain medium can be modeled macroscopically by a complex permittivity $\epsilon_{b}=\epsilon^{'}_{b}+i\epsilon^{''}_{b}$ where the real part $\epsilon^{'}_{b}$ has contributions from (i) the host and (ii) the real-valued permittivity induced by $\epsilon^{''}_{b}$. Similarly the complex permittivity of the metal is $\epsilon_{m}=\epsilon^{'}_{m}+i\epsilon^{''}_{m}$ with $\epsilon^{'}_{m} <0$ and $|\epsilon^{'}_{m}| \gg \epsilon^{''}_{m}, \epsilon^{'}_{b},|\epsilon^{''}_{b}|$. In the closed $\Xi-$configuration, the transition $|c\rangle \leftrightarrow |a\rangle$ and $|a\rangle \leftrightarrow |b\rangle$ of the quantum emitters are driven by the control and probe field of Rabi frequency $\Omega_{a}$ and $\Omega_{b}$ respectively. 

The Hamiltonian in the interaction picture can be written as 
\begin{equation}\label{eq2}
\mathscr{H}=-\left(\Omega_{b}e^{i\Delta_{b}t}\left |a \rangle \langle b \right |+\Omega_{a}e^{i\Delta_{a}t}\left |c \rangle \langle a \right |\right)+\text{H.c}.
\end{equation} 
Here the detunings are defined as $\Delta_{a}=\omega_{ca}-\omega_{c}$, $\Delta_{b}(\omega)=\omega_{ab}-\omega$. The decay rates from $|a\rangle \rightarrow |b\rangle$ is $\gamma_{b}$, $|c\rangle \rightarrow |a\rangle$ is $\gamma_{a}$ and $|c\rangle \rightarrow |b\rangle$ is $\gamma_{c}$ while the incoherent pumping rate from $|b\rangle \rightarrow |c\rangle$ is $g$. Incorporating these rates, the equation of motion for the density matrix $\varrho$ satisfies the Liouville-von Neumann equation
\begin{equation}\label{eq3}
\begin{split}
\dot{\varrho}=-\frac{i}{\hbar}[\mathscr{H},\rho]-\frac{1}{2}\{\Gamma,\varrho\},
\end{split}
\end{equation}
where $\{\Gamma,\varrho\}= \Gamma\varrho +\varrho\Gamma$. Furthermore we assume that the driving field is strong enough that it does not change significantly with time and can be considered as constant. Any transition $\varrho_{ij}$ exhibit gain or absorption is determined by the imaginary part of $\varrho_{ij}$. Conventionally if Im[$\varrho_{ij}]<0(>0)$ then the transition $|i\rangle\leftrightarrow |j\rangle$ exhibits gain(absorption). The linear response of the three-level quantum emitter based gain medium (see Fig. 1) is given by the dielectric function $\epsilon_{b}=\epsilon^{'}_{b}+\chi^{(1)}_{b}(\omega)$ where $\epsilon^{'}_{b}$ is the dielectric constant of the host and the drive field dependent susceptibility\cite{ScullyZub}
\begin{equation}\label{eq4}
\chi^{(1)}_{b}(\omega)\simeq -i\frac{N|\wp_{ab}|^{2}}{\epsilon_{0}\hbar}\left\{\frac{n^{(0)}_{ab}\Gamma_{cb}\Gamma_{ca}+n^{(0)}_{ca}\Omega_{a}^{2}}{(\Gamma_{cb}\Gamma_{ab}+\Omega_{a}^2)\Gamma_{ca}}\right\},
\end{equation}
where the coherence relaxation terms are given as
\begin{equation} 
\begin{split}
\Gamma_{ab}=&\gamma_{ab}+i\Delta_{b}(\omega),\quad \Gamma_{ca}=\gamma_{ca}+i\Delta_{a},\\
& \Gamma_{cb}=\gamma_{cb}+i(\Delta_{a}+\Delta_{b}(\omega)).
\end{split} 
\end{equation} 
and $\gamma_{ij}$ has the form 
\begin{equation}
\begin{split}
\gamma_{ab}=(\gamma_{b}+g)/2+&\gamma^{(ab)}_{d}, \gamma_{ca}=(\gamma_{a}+\gamma_{b}+\gamma_{c})/2+\gamma^{(ca)}_{d},\\
 \gamma_{cb}&=(\gamma_{a}+\gamma_{c}+g)/2+\gamma^{(cb)}_{d}.
 \end{split}
 \end{equation}
Here $\gamma^{(ij)}_{d}$ is the phase relaxation (or dephasing) rate of the coherence $\varrho_{ij}$ due coupling with phonons, surface defects etc\cite{Woggon}. The population inversion is defined as $n^{(0)}_{ij}=\varrho^{(0)}_{ii}-\varrho^{(0)}_{jj}$. At resonance i.e $\Delta_{b}=0, \Delta_{a}=0$,  $\chi^{(1)}_{b}(\omega_{ab})$ is purely imaginary, thus the complex part of the permittivity of the gain medium is $\epsilon^{''}_{b}=\text{Im}[\chi^{(1)}_{b}(\omega_{ab})]$. Now we can rewrite the dispersion relation Eq.(\ref{eq1}) with the control parameter $\Omega_{a}$ as
\begin{equation}\label{eq04}
\begin{split}
k^{2}_{x}(\Omega_{a})=k^{2}_{0}\left[\frac{\epsilon_{b}(\Omega_{a})\epsilon_{m}}{\epsilon_{b}(\Omega_{a})+\epsilon_{m}}\right].
\end{split}
\end{equation}
We solve Eqs.(\ref{eq3}) for the steady-state population inversion on the transition $|a\rangle\leftrightarrow |b\rangle$ and obtain
\begin{equation}\label{eq5}
\begin{split}
n^{(0)}_{ab}=\frac{[g\gamma_{a}-\gamma_{b}(\gamma_{a}+\gamma_{c})]\Gamma_{ca}+2(g-\gamma_{b}-\gamma_{c})\Omega^{2}_{a}}{[\gamma_{b}(\gamma_{a}+\gamma_{c})+g(\gamma_{a}+\gamma_{b})]\Gamma_{ca}+2(2g+\gamma_{b}+\gamma_{c})\Omega^{2}_{a}}.
\end{split}
\end{equation} 
From Eq.(\ref{eq1}), one can easily obtain the threshold value of the imaginary part of permittivity of the gain medium $\epsilon^{''}_{b, th}$ for lossless propagation of SPPs. In the limit $\epsilon^{''}_{m}, \epsilon^{'}_{b} \ll |\epsilon^{'}_{m}|$,  the threshold value is given as\cite{Plotz69} $\epsilon^{''}_{b,\,th}=-\epsilon^{''}_{m}\epsilon^{'2}_{b}/\epsilon^{'2}_{m}$ which can be achieved in experiments for some combination of the parameters.  In the absence of a coherent drive ($\Omega_{a}=0$) positive gain requires $n^{(0)}_{ab}>0$ i.e population inversion on the transition $|a\rangle \leftrightarrow |b\rangle$ which gives the threshold value of the pump rate, using Eq.(\ref{eq5}) as 
\begin{equation}
g_{th}=\gamma_{b}\left(1+\frac{\gamma_{c}}{\gamma_{a}}\right).
\end{equation}
Generally $\gamma_{b} \gg \gamma_{a}, \gamma_{c}$ and assuming $\gamma_{c}/\gamma_{a} \simeq 0.1$-$0.2$, the rate of incoherent pump $g$ should exceed the decay rate $\gamma_{b}$. From Eq.(\ref{eq4}) we obtain 
\begin{equation}\label{eq6}
\epsilon^{''}_{b}\simeq -\xi\gamma_{a}\left(\frac{1}{\gamma_{b}}-\frac{1}{g}\right)\left(\frac{\gamma_{b}+g}{2}+\gamma_{d}\right)^{-1},
\end{equation}
where the constant $\xi=N|\wp_{ab}|^{2}/\epsilon_{0}\hbar$. Let us now define $g_{c}$ as the critical incoherent pump rate such that the Im$[k_{x}]=0$ i.e $\epsilon^{''}_{b} \rightarrow \epsilon^{''}_{b,\,th}$. The value of $\epsilon^{''}_{b,\,th}$ is determined by the material properties of the metal $\epsilon^{'}_{m}, \epsilon^{''}_{m}$ and the dielectric host $\epsilon^{'}_{b}$(assuming real-valued permittivity induced by $\epsilon^{''}_{b}$ is zero). We can calculate $g_{c}$ as the root of the following equation obtained from Eq.(\ref{eq6})
\begin{equation}\label{eq7}
g^{2}+(\gamma_{b}+2\gamma_{d}-2\theta/\gamma_{b})g+2\theta=0,
\end{equation}
where $\theta=-\xi\gamma_{a}/\epsilon^{''}_{b,\,th}$ and the critical value is given as
\begin{equation}
2g_{c}=-\alpha\pm\sqrt{\alpha^{2}-8\theta}.
\end{equation}
Here we have defined $\alpha=\gamma_{b}+2\gamma_{d}-2\theta/\gamma_{b}$. Using simple algebra we see that in order to have at least one positive critical pump rate $g_{c}$ we require $\alpha <0$ and $\alpha^{2}-8\theta \ge 0$. Till now we have discussed about propagation, we also know that the imaginary part of $k_{z,\alpha}$ characterizes the confinement of SPP in the metal/phaseonium regions. Using Eq.(\ref{eq1}), the condition that Im[$k_{z,b}]>0$ requires\cite{Nez04} $\epsilon^{''2}_{b}+\epsilon^{''}_{b}\epsilon^{''}_{m}+2\epsilon^{'}_{b}(\epsilon^{'}_{b}+\epsilon^{'}_{m})<0$. This gives the range of the gain(absorption) for which SPP will be confined in the phaseonium region of the interface. The knowledge of this range of $\epsilon^{''}_{b,l}<\epsilon^{''}_{b}<\epsilon^{''}_{b,u}$ is crucial  to keep the SPPs confined even with the extra boost in gain provided via quantum coherence.

\begin{figure}[b]
\centerline{\includegraphics[height=5.2cm,width=0.5\textwidth,angle=0]{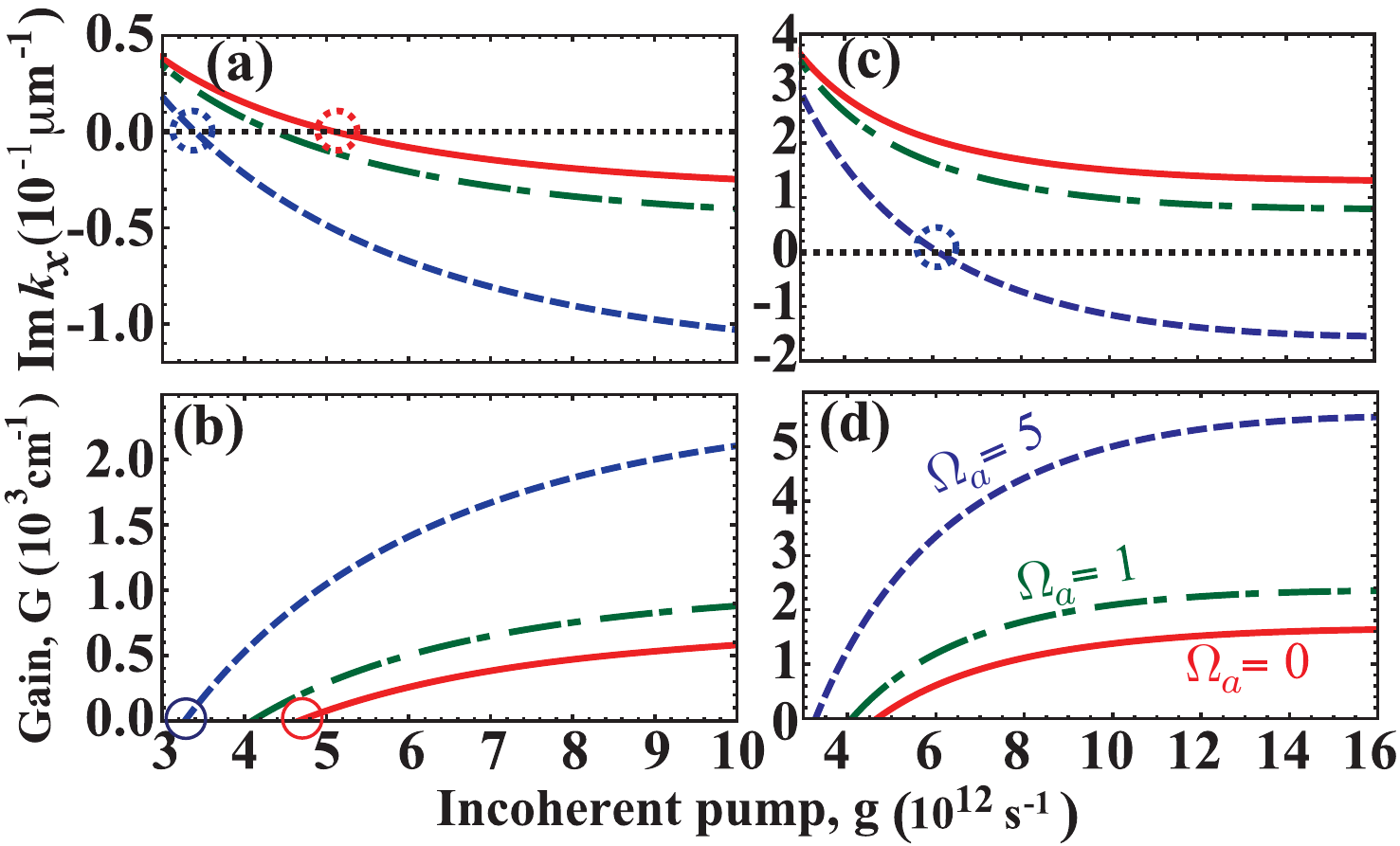}}
\caption{Plot of Im$k_{x}$ as a function of the incoherent pumping rate $g$ for different driving fields: $\Omega_{a}=0$ (solid red), $1\times 10^{12}\text{s}^{-1}$ (dashed dot green) and $5\times 10^{12}\text{s}^{-1}$(dashed blue). The dotted circles corresponds to lossless propagation. Here (a,b) are for telecommunication (@1550nm) while (c,d) corresponds to visible (@653nm) wavelengths excitation of the SPP. We have modeled the gain medium using the following realistic parameters: $\gamma_{b}=4\times 10^{12}\text{s}^{-1}$, $\gamma_{a}=6\times 10^{11}\text{s}^{-1}$, $\gamma_{c}=1\times 10^{11}\text{s}^{-1}$, $|\wp_{ab}|=8\times 10^{-30}$Cm and $N=10^{26}\text{m}^{-3}$. The dephasing of coherence $\varrho_{ab}$ is $\gamma_{ab} \simeq 1.4\times 10^{13}\text{s}^{-1}$ while for the other transitions $(|i\rangle\leftrightarrow |j\rangle)$ we considered $\gamma_{ij} \simeq 4\times 10^{12}\text{s}^{-1}$. The solid circles in Fig 2(b) corresponds to the threshold value $g_{th}$ for gain. Coherent drive induces higher gain which assists reducing the threshold pump for gain(shown by solid circles in (b)) and lowers the critical pump required for lossless propagation (shown by the dotted circles in (a) and (c)). }
\label{SPPFig2}
\end{figure}
Next we present the numerical simulation results obtained by solving the system of Eqs.(\ref{eq3}). In our model the lower transition ($|a\rangle \leftrightarrow |b\rangle$) of the three-level gain medium is resonantly coupled to the SP mode of the metal.  We also assumed that the driving field, which is resonantly coupled with the transition ($|c\rangle \leftrightarrow |a\rangle$), does not excite any SP modes. This assumption circumvents any heating effect due to the drive field.  We will consider propagation of a SPP launched along the MP interface in two wavelength regimes (a) telecommunication @1550nm and (b) visible @653nm. The metal portion consists of  gold plate with dielectric functions\cite{Palik} $\epsilon_{m}=-131.948 +12.65i$ and $-9.895+1.0458i$ at the telecommunication and visible wavelengths respectively. The dielectric constant of the host is assumed to be $\epsilon^{'}_{b} = 2.25$.  In the close proximity of metal surfaces, the decay rates of the emitter can be enhanced up to 2-3 orders of magnitudes with respect to free-space decay rate\cite{Larkin04}. These enhancement can be described by the combined effect of variations in local density of states(LDS) and excitation of SPs, lossy surface waves(LSW)\cite{Ford84,Barnes98} etc. For an optimal position of the emitters from the interface an efficient coupling to the SP modes can be achieved\cite{Ford84,Barnes98, Lukin07}. Taking these factors into account along with inhomogeneous broadening due to the host dielectric,  we have modeled our three-level emitter based gain medium with parameters given in Fig. 2.

Fig. 2(a,c), obtained by solving the system of Eqs. (\ref{eq3}) for different values of the Rabi frequency $\Omega_{a}$, shows the imaginary part of the propagation vector $k_{x}$ in the presence of the coherent drive at telecommunication and visible wavelengths respectively. While in the visible regime propagation of the SP is lossy, on the other hand the gain requirement in the telecom regime is not high and thus lossless propagation can be easily achieved even in the absence of external drive, as shown by dotted circles in Fig. 2(a). In the telecom regime, the coherent drive enhances gain at a given incoherent pump $g$ and reduce the critical value $g_{c}$ for lossless propagation by $30\%$ with a drive $\Omega_{a}=5\times 10^{12}\text{s}^{-1}$. 

The reduction in the critical value $g_{c}$ follows the enhancement of the optical gain power defined as\cite{Plotz69} $G=-k_{0}\epsilon^{''}_{b}/(\epsilon'_{b})^{1/2}$. The gain enhancement is due to the quantum coherence induced in the three-level gain medium by the drive $\Omega_{a}$, which also reduces the threshold pump ($g_{th}$) to observe gain on the transition $|a\rangle \leftrightarrow |b\rangle$ as shown by solid circles in Fig. 2(b). At moderate drive $\Omega_{a}=5\times 10^{12}\text{s}^{-1}$ we observed a $30\%$ reduction in $g_{th}$. In Figs. 2(b,d) we have shown the plot of optical gain power for different values of the Rabi frequency $\Omega_{a}$ which clearly shows gain enhancement. Thus coherent drive relaxes the conditions required for lossless propagation, allows reduction in the critical $(g_{c})$ and the threshold $(g_{th})$ pump values. We calculated the upper value of gain $G$ for which the SPPs are confined in phaseonium region as $2.25\times 10^{4}\text{cm}^{-1}$ and $5.39\times 10^{4}\text{cm}^{-1}$ for visible and telecommunication wavelengths respectively. These gain values are well above the gain we have considered here, thus even in the presence of coherent drive SPPs remains confined metal/phaseonium regions of the interface.

\begin{figure}[t]
\centerline{\includegraphics[height=5.3cm,width=0.47\textwidth,angle=0]{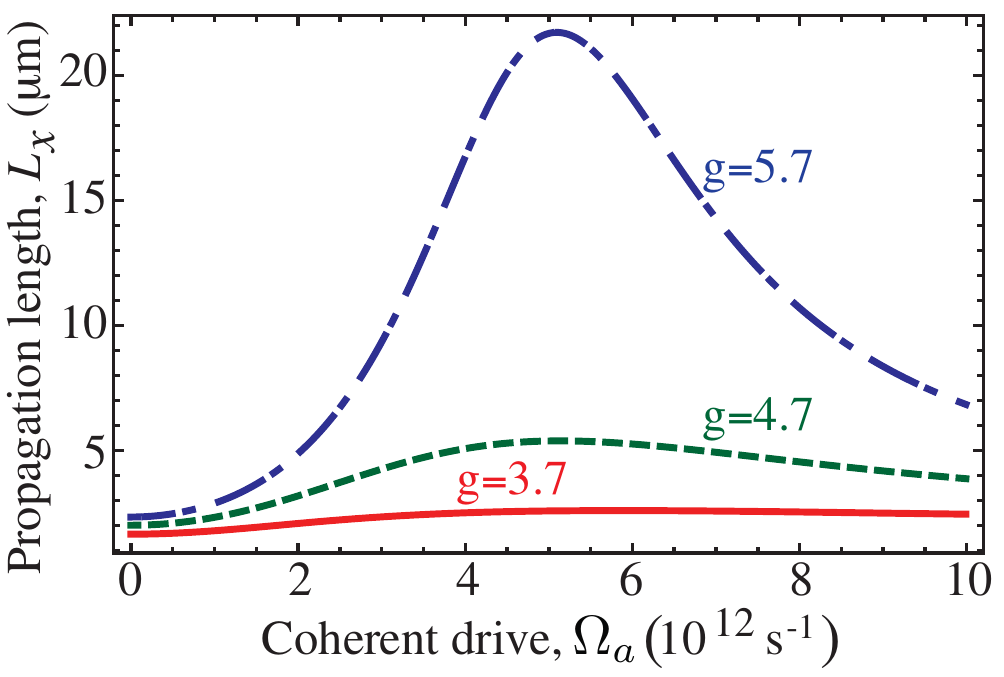}}
\caption{Plot of the propagation length $L_{x}$ of the SPP excited at visible wavelength (@653nm) as a function coherent drive Rabi frequency $\Omega_{a}$ for three different values of $g= $3.7$\times 10^{12}\text{s}^{-1}$(solid red), 4.7$\times 10^{12}\text{s}^{-1}$(dashed green), 5.7$\times 10^{12}\text{s}^{-1}$(dashed dot blue) respectively. All other parameters are same as Fig.(2). Coherent drive enhances the propagation length substantially for optimum value of $\Omega_{a}$.}
\label{SPPFig3}
\end{figure}
Next we demonstrate how the propagation length can be controlled by the coherent drive when the propagation of SPPs are lossy. Here we considered the visible regime at several pumping rates for which lossless propagation cannot be achieved even in the presence of the drive field. Fig. 3 shows the simulation result on propagation length $L_{x}=1/2\text{Im}k_{x}$ as a function of drive Rabi frequency $\Omega_{a}$ for three choices of the incoherent pump rate $g=3.7, 4.7, 5.7 \times 10^{12}s^{-1} $. At low incoherent pump, the gain enhancement due to coherent drive is marginal and thus the effect on propagation is not pronounced. However at higher pump rate i.e higher gain without drive, the additional gain using coherent drive is substantial to enhance the propagation length.  At $g=5.7\times 10^{12}\text{s}^{-1}$, without drive we calculated the propagation length $L_{x}\sim 2.33\mu m$ which is enhanced by an order of magnitude to $L_{x} \simeq 21.72\mu m$ with a drive $\Omega_{a}\simeq 5\times 10^{12}\text{s}^{-1}$. These results clearly demonstrate controllable coherence-enhanced propagation of SPP using an external source. 

In summary, we have theoretically demonstrated quantum coherence-enhanced propagation of SPPs along the MP interface. We demonstrate lossless propagation at visible wavelength along with lower pumping requirements at both visible and telecommunication wavelengths. Indeed quantum coherence can be used as a boost when (effective) two-level system based gain medium are insufficient or require high pumping requirements. Other approaches to enhance SPP propagation lengths like using buried metal grating\cite{Jose11}, chemisorption\cite{Gavri10}, via coupling to asymmetric waveguide structures\cite{Mont08} etc. have been studied recently.  The interesting feature of the present approach is that it provides an external control parameter for the propagation of SPP over long range if not lossless. Such optical control\cite{Kevin09, Samson11,Dorfman13} of SPs and SPPs holds promise for quantum control and add a new dimension to the field of nanoplasmonics.

P. K. Jha would like to thank Y. V. Rostovtsev, A. Salandrino, C. Wu and H. Suchowski for fruitful discussions.

\end{document}